\documentclass[11pt]{article}

\usepackage{fullpage}
\usepackage{amssymb,amsmath}
\usepackage{graphicx, epsfig}

\usepackage[compact]{titlesec}

\def\idtt#1{\ensuremath{\mathtt{#1}}}

\newtheorem{theorem}{Theorem}
\newtheorem{lemma}{Lemma}

\newtheorem{property}{Property}

\newenvironment{proof}{\trivlist\item[]\emph{Proof}:}%
{\unskip\nobreak\hskip 1em plus 1fil\nobreak$\Box$
\parfillskip=0pt%
\endtrivlist}

\newenvironment{proofsk}{\trivlist\item[]\emph{Proof Sketch}:}%
{\unskip\nobreak\hskip 1em plus 1fil\nobreak$\Box$
\parfillskip=0pt%
\endtrivlist}

\newenvironment{itemize*}%
  {\begin{itemize}%
    \setlength{\itemsep}{0pt}%
    \setlength{\parskip}{0pt}%
    \setlength{\parsep}{0pt}%
    \setlength{\topsep}{0pt}%
    \setlength{\partopsep}{0pt}%
  }%
  {\end{itemize}}%

\newcommand{\cB}{{\cal B}}

\newcommand{\oS}{\overline{S}}
\newcommand{\oAC}{\overline{AC}}
\newcommand{\tAC}{\widetilde{AC}}

\newcommand{\sfind}{\idtt{find}}
\newcommand{\sinsert}{\idtt{insert}}
\newcommand{\sinsertaux}{\idtt{insertAux}}
\newcommand{\sdelete}{\idtt{delete}}
\newcommand{\sdeleteaux}{\idtt{deleteAux}}
\newcommand{\sdeleteauxlong}{\idtt{deleteAux1}}

\newcommand{\spred}{\idtt{pred}}
\newcommand{\ssucc}{\idtt{succ}}
\newcommand{\sset}{\idtt{set}}
\newcommand{\smid}{\idtt{mid}}
\newcommand{\sleft}{\idtt{left}}
\newcommand{\sleftmax}{\idtt{leftmax}}
\newcommand{\mlab}{\idtt{lab}}
\newcommand{\sright}{\idtt{right}}
\newcommand{\srightmin}{\idtt{rightmin}}
\newcommand{\sgroup}{\idtt{group}}
\newcommand{\parent}{\idtt{parent}}

\newcommand{\mins}{\idtt{minsert}}
\newcommand{\mdel}{\idtt{mdelete}}
\newcommand{\eps}{\varepsilon}

\begin{document}

\title{\vspace*{3cm} Searching in Dynamic Catalogs on a Tree}
\author{Yakov Nekrich\thanks{Department of Computer Science, 
 University of Bonn. 
 Email {\tt yasha@cs.uni-bonn.de}}
}
\date{}
\maketitle
\begin{abstract}
In this paper we consider the following  modification of the iterative search
 problem.
We are given a tree $T$, so that a dynamic  catalog $C(v)$ is associated with 
every tree node $v$. 
For any $x$ and for any node-to-root path $\pi$ in $T$, we must find the predecessor 
of $x$ in $\cup_{v\in \pi} C(v)$.
We present a linear space dynamic data structure that supports such 
queries in $O(t(n)+|\pi|)$ time, where $t(n)$ is the time needed to search in 
one catalog and $|\pi|$ denotes the number of nodes on path $\pi$.

We also consider the reporting variant of this problem, in which for any 
$x_1$, $x_2$ and for any path $\pi'$, 
all elements of $\cup_{v\in \pi'} (C(v)\cap [x_1,x_2])$ must be reported;
here $\pi'$ denotes a path between an arbitrary node $v_0$ and its
 ancestor $v_1$. 
We show that such queries can be answered in $O(t(n)+|\pi'|+ k)$ time, 
where $k$ is the number of elements in the answer.

To illustrate applications of our technique, we describe the first dynamic 
data structures for 
the stabbing-max problem, the horizontal point location problem, 
and the orthogonal line-segment intersection problem  
with optimal $O(\log n/\log \log n)$ query time and poly-logarithmic 
update time. 
\end{abstract}
\setcounter{page}{0}
\thispagestyle{empty}
\clearpage
\section{Introduction}
The situation when we must search for  the position of a value $x$ in many 
ordered sets frequently arises in data structures and computational geometry 
problems. 
The brute force approach of searching for $x$ in every  set ``from scratch'' 
can be improved
if there are restrictions on the order in which the sets can be  searched.
Such improvements for some important problems were suggested by several 
researchers, see e.g.,~\cite{VW82,W88}. Chazelle and Guibas described in 
their seminal paper~\cite{CG86} a general 
data structuring technique, called fractional cascading, that addresses the 
general problem of searching in multiple sets. 
The fractional cascading technique 
solves the following iterative search problem: 
We are given a graph $G$, called the catalog graph,
so that an ordered set $C(v)\subset U$, called a catalog, 
 is associated with every graph node. A query consists of an element 
$x\in U$ and a subgraph $G'$ of $G$. The goal is to find the predecessor\footnote{The predecessor of $x$ in $S$, denoted $\spred(x,S)$, is the
 largest $e\in S$ such that $e\leq x$; the successor of $x$ in $S$, 
denoted 
$\ssucc(x,S)$, is the smallest $e\in S$ such that $e\geq x$.} 
of $x$ in \emph{each} catalog $C(v)$ for $v\in G'$. 
In this paper we consider the following modification of the iterative search, 
further called \emph{multiple catalog searching} problem:  
the graph $G$ is a rooted tree, the subgraph $G'$ is a node-to-root 
path $\pi$, and we must search in \emph{the union of all} catalogs $C(v)$, 
$v\in \pi$. We also consider the reporting variant, further called 
multiple catalog reporting,  in which all 
elements $e\in C(v)$, $v\in \pi$, that belong to the query range 
$[x_1,x_2]$ must be reported. 

Although the problems addressed in this paper are more restrictive 
than iterative searching, they can be applied in many situations in 
which iterative searching is traditionally used. 
We show that multiple catalog searching and reporting queries 
can be answered by spending constant time in each node $v$ of $\pi$ 
if $\pi$ is sufficiently large (ignoring the time to output all elements 
in the answer).
This enables  us to obtain for the first time 
dynamic data structures with optimal query time and poly-logarithmic 
update time for 
point location in a set of horizontal segments, 
stabbing-max, and orthogonal line-segment intersection 
reporting. 
\\{\bf Previous and Related Work.}
Chazelle and Guibas~\cite{CG86} 
showed that it is 
possible to identify the predecessor of $x$ in $C(v)$ for \emph{each} 
catalog $C(v)$, $v\in G'$, in $O(t(n)+|G'|)$ time, where $n$ denotes the 
total number of elements in all catalogs, $|G'|$ is the number of nodes in 
$G'$, and $t(n)$ is the time needed to search in one catalog. 
The dynamic version of the fractional cascading is considered by Mehlhorn 
and N\"aher~\cite{MN90}; in~\cite{MN90} the authors described how to support
insertions into and deletions from a catalog $C(v)$ in $O(\log \log n)$ 
time if a pointer to  the deleted element $x$ or the predecessor of an 
inserted element $x$ is given; the data structure of~\cite{MN90} supports 
queries in  $O(t(n)+|G'|\log \log n)$ time, i.e., the search takes 
$O(\log \log n)$ time in  each node of $G'$. 
Imai and Asano~\cite{IA87} considered the semi-dynamic scenario, when new 
elements can be inserted but deletions are not supported. The result 
of ~\cite{IA87} can be used to support insertions in $O(\log^* n)$ time 
and search in $O(t(n)+|G'|\log^* n)$ time in the pointer machine 
model~\cite{T79};
another result of~\cite{IA87} can be used to support insertions in 
$O(1)$ time and search in $O(t(n)+|G'|)$ time in the RAM model. 
Since~\cite{IA87,MN90}, the dynamic fractional cascading was applied 
to a number of data structure problems, e.g., point location, 
range reporting, and segment intersection. The technique 
was also extended  e.g., to support iterative search in graphs 
with super-constant local degree~\cite{R92} and to  the case when elements
stored in different catalogs belong to different ordered sets,
 e.g.~\cite{BJM94,ABG06}.
However, there is no currently known dynamic data structure that 
supports iterative search in $o(\log \log n)$ time per catalog 
(ignoring the $O(t(n))$  term).  Since fractional cascading relies 
on the union-split-find queries, and union-split-find queries 
cannot be answered in $o(\log \log n)$ time~\cite{MNA88}, 
it appears that we must spend $\Omega(\log \log n)$ time in each node 
to solve the iterative searching problem. 
\\{\bf Our Results.}
The fractional cascading ~\cite{CG86} technique and 
its variants for the dynamic and semi-dynamic scenarios~\cite{MN90,IA87} 
can be applied when the catalog graph is an arbitrary graph with locally 
bounded degree (e.g., any graph with bounded degree; see~\cite{CG86} for
 precise definition). 
In our scenario the catalog graph is a rooted tree and all catalogs 
$C(v)$ for all nodes $v$ on the path $\pi$ must be searched. 
Moreover, instead of searching for $x$ in \emph{each} catalog, we search 
in \emph{all} catalogs. That is, the query consists of a value $x$ and 
a path $\pi$ between a node $u$ and the root of the tree; the answer to 
the query 
is the predecessor $p_x$ of $x$ in the union of all catalogs on $\pi$,
$p_x=\spred(x,\cup_{v\in\pi} C(v))$. Henceforth, such queries will be called
\emph{multiple catalog searching} queries. 
We obtain the following results with a linear space data structure:\\ 
1. 
Multiple catalog searching queries   can be answered in 
$O(t(n)+(1/\eps)|\pi|)$ time, and updates are supported in 
$O(\log^{\eps} n)$ time for any $\eps>0$.\\ 
2. 
Multiple catalog searching queries   can be answered in 
$O(t(n)+|\pi|\log \log \log n)$ time, and 
updates are supported in $O(\log \log n)$ time. \\
 Other trade-offs between query and update times are
 described in Theorem~\ref{theor:mcs}. We assume that a pointer 
to the position of an inserted or deleted element in the data structure 
is known for the update operation.

We also consider the multiple catalog reporting problem. A query 
consists of values $x_1,x_2\in U$ and a path $\pi$ from a node $v_0$ 
to a node $v_1$, such that $v_1$ is the ancestor of $v_0$. 
The answer to the query consists 
of all elements $e\in \cup_{v\in \pi} C(v)$, such that 
$x_1\leq e \leq x_2$.\\ 
1. 
Multiple catalog reporting queries   can be answered in 
$O(t(n)+ (1/\eps)|\pi| + k)$ time, where $k$ is the number of 
elements in the answer, and updates are supported in 
$O(\log^{\eps} n)$ time for any $\eps>0$. \\
2. 
Multiple catalog reporting queries   can be answered in 
$O(t(n)+|\pi|\log\log \log n + k)$ time, where $k$ is the number of 
elements in the answer, and updates are supported in $O(\log \log n)$ time. 
\\ Again, the space usage of our data structure is linear in the total number
 of elements in all catalogs. Other trade-offs between query and update times are described 
in Theorem~\ref{theor:mcrep}. 
Dynamic range reporting in a single catalog was considered 
in~\cite{MPP05,M06}.
The data structure of~\cite{MPP05, M06} supports queries 
and updates in $O(\log \log \log U)$ and $O(\log \log U)$ time 
respectively, where $U$ is the size of the universe.  
Another variant of their data structure supports 
queries in $O(1)$ time and updates in $O(\log^{\eps} U)$ time. 
Besides that, the data structure described in~\cite{MPP05} uses randomization 
and relies  on a more extensive set of basic arithmetic operations.

Finally, we consider the multiple catalog maxima problem.
A query consists of a path $\pi$ from a node $v_0$ 
to a node $v_1$, such that $v_1$ is the ancestor of $v_0$; we must output 
the maximal element in every catalog $C(v)$, $v\in\pi$.  
For a tree with node degree $O(\log^{1/4}n)$ such queries can be answered 
in $O(|\pi|)$ time. Insertions and deletions are supported in $O(\log\log n)$ 
and $O((\log\log n)^2)$ time respectively. Moreover, in this case we  extend 
the definition of update operations, so that an element can be 
 simultaneously inserted into (deleted 
from) any catalogs $C(v_f),\ldots,C(v_l)$ where $v_f,\ldots,v_l$ are sibling 
nodes. This result, described in section~\ref{sec:mcmax},  is obtained with 
a different, simpler technique.
\\{\bf Applications.}
As an illustration of our technique, 
we present  dynamic data structures for several problems 
that for the first time achieve $O(\log n/\log\log n)$  query time in the
 word RAM model. 
The marked ancestor problem~\cite{AHR98} can be reduced to each of the problems 
described below, see~\cite{AHR98}. In~\cite{AHR98}, the authors also show
 that any data structure with poly-logarithmic update time and 
poly-logarithmic word size needs $\Omega(\log n/\log \log n)$ time to 
answer the marked ancestor problem.  
Hence, we obtain data structures 
with optimal query time for all  considered problems.    
\\{\bf Horizontal Point Location.}
In the horizontal point location problem aka vertical ray shooting problem, 
a set of $n$ horizontal segments is stored in the 
data structure, so that for a query point $q$ the segment immediately 
below (or immediately above) $q$ can be reported. 
Giyora and Kaplan~\cite{GK09} describe a linear space RAM data structure with 
$O(\log n)$ query and update times in the RAM model. 
We refer to~\cite{GK09} for a detailed  description of previous results.
Although the $O(\log n)$ time is optimal if we can manipulate segments
by comparing their coordinates, the query time can be improved 
in the word RAM model.  
In this paper we present a data structure that supports queries in 
 $O(\log n/\log \log n)$ time and updates in $O(\log^{1+\eps}n)$ 
amortized time; our data structure uses $O(n)$ space. 
As explained above, this query time is optimal. 
\\{\bf Retroactive Searching.}
In the retroactive searching problem, introduced by Demaine et.al.~\cite{D04},
the data structure maintains a sequence of keys. 
Each key can be inserted at time $t_I$ and deleted at time $t_D>t_I$. 
The answer to a query $(q,t)$ is the element that precedes $q$ at time $t$. 
It was shown in~\cite{GK09} that retroactive searching is equivalent 
to the horizontal point location problem. Thus our result for horizontal 
point location demonstrates
that retroactive searching queries can be answered in  $O(\log n/\log \log n)$ 
time in the word RAM model. 
\\{\bf Stabbing-Max Data Structure.}
In the stabbing-max problem, we maintain a set of axis-parallel 
$d$-dimensional  rectangles, and each rectangle $s$
has priority $p_s$. Given a query point $q$, the stabbing-max data 
structure finds a rectangle with maximum priority that contains $q$. 
The one-dimensional data structure of Kaplan, Molad, and Tarjan~\cite{KMT03}
supports queries and insertions in $O(\log n)$ time, deletions 
in $O(\log n\log \log n)$ time, and uses $O(n)$ space. 
The data structure of Agarwal, Arge, and Yi~\cite{AAY05} 
also uses linear space and supports queries and updates in $O(\log n)$ time. 
See~\cite{KMT03,AAY05, Th03} for a more extensive description of 
previous results. 

In this paper we describe two data structures that support one-dimensional 
stabbing-max queries in optimal $O(\log n/\log \log n)$ time.
The first data structure uses $O(n\log n/\log \log n)$ space 
and supports insertions and deletions in $O(\log n)$ and  $O(\log n\log\log n)$ time respectively. The second  data structure uses $O(n)$ space but supports 
updates in $O(\log^{1+\eps}n)$ time.
\\{\bf Orthogonal Line-Segment Intersection.}
In this   problem a set of horizontal 
segments is stored in a data structure, so that for a vertical query segment 
$s_v$ all segments that intersect with $s_v$ can be reported. 
The data structure of Cheng and Janardan~\cite{CJ90} supports such queries 
in $O(\log^2 n+k)$, where $k$ is the number of segments in the answer. 
Mehlhorn and N\"aher reduced the query time to 
$O(\log n \log \log n +k )$ using dynamic fractional cascading. 
The fastest previously known data structure of Mortensen~\cite{M03} 
supports queries and updates in $O(\log n+k)$  and $O(\log n)$ time 
respectively and uses $O(n\log n/\log \log n)$ space. 
In this paper we present a $O(n\log n /\log \log n)$ space data structure 
that answers queries in $O(\log n/\log \log n + k)$ time and supports updates 
in $O(\log^{1+\eps}n)$ time. 

All results presented in this paper are valid in the word RAM computation
 model. We assume that 
every element (resp. every point) fits into one machine word and that 
additions, subtractions, and bit operations can be performed in constant
time. We also assume that the most significant bit (MSB) of an integer 
can be found in $O(1)$ time. It is possible to find MSB in $O(1)$ time using 
$AC^0$ operations~\cite{AMT99}.
Throughout this paper $\eps$ denotes an arbitrarily small positive 
constant.

\section{Main Idea}
\label{sec:overview}
In this section we sketch the main ideas of our approach.
We start by showing how the fractional cascading technique can be 
used to solve the multiple catalog searching problem.
Then, we describe the difference between the fractional cascading 
and our approach.

We construct\footnote{We describe a simplified version of the fractional 
cascading technique because we are only interested in searching catalogs that 
lie on a node-to-root path.} 
augmented catalogs $AC(v)\supseteq C(v)$ for all nodes 
$v$ of $T$ starting at the root. For the root $v_R$, $AC(v_R)=C(v_R)$. 
If $AC(u)$ for a node $u$ is already constructed, then 
$AC(u_j)$ for a child $u_j$ of $u$ consists of some elements from $AC(u)$ and 
all elements of $C(u_j)$. 
Elements of $C(u)$ and $AC(u)\setminus C(u)$ are called proper elements 
and improper elements respectively. For every improper element $e\in AC(u)$, 
there is a copy $e'$ of $e$ that is stored in $AC(\parent(u))$. 
Elements $e$ and $e'$ are provided with pointers to each other and are 
called a \emph{bridge}.

We want to organize the search procedure 
in such way that only a small number of elements  in 
every visited node must be examined. Using fractional cascading, we  can
 guarantee    
that there are $O(d)$ elements of $AC(w)$ between any two 
improper elements of $AC(v)$,
where $v$ is any node of $T$ except of the root and $w$ is the parent of $v$.
Each element stored in the augmented catalog $AC(v)$ belongs either 
to $C(v)$ or to a catalog $C(w)$ for some ancestor $w$ of $v$. 
Hence, $\cup_{v\in \pi} AC(v)=\cup_{v\in \pi} C(v)$ for any node-to-root path 
$\pi=v_0,v_1,\ldots,v_R$. 
Therefore  multiple catalog searching (unlike general iterative searching) 
is equivalent to finding the predecessor in  $\cup_{v\in \pi} AC(v)$.
This suggests the following method for 
searching in $\cup_{v\in \pi} C(v)$:   
The search procedure starts by identifying  $p(v_0)=\spred(x,AC(v_0))$ and 
$s(v_0)=\ssucc(x,AC(v_0))$. For every node $v_i$, $i>0$, 
we find $p(v_i)=\spred(x,AC(v_0)\cup\ldots\cup AC(v_i))$ and 
$s(v_i)=\ssucc(x,AC(v_0)\cup\ldots\cup AC(v_i))$. 
Suppose that $p(v_i)$ and  $s(v_i)$ for some $i\geq 0$ are known. 
To identify $p(v_{i+1})$ and $s(v_{i+1})$, we only need to examine 
elements of $AC(v_{i+1})$ that belong to the interval $[p(v_i),s(v_i)]$.
Since there is no element $e\in AC(v_i)$ between $p(v_i)$ and $s(v_i)$, 
there are $O(d)$ elements of $AC(v_{i+1})$ between $p(v_i)$ and $s(v_i)$, 
where $d$ is the maximal node degree of $T$.  
For $d=\log^{O(1)}n$, we can search in a set of $O(d)$ elements in $O(1)$ 
time~\cite{FW94}.

The only issue is how to quickly find elements of $AC(v_{i+1})$ 
that are between $p(v_i)$ and $s(v_i)$.  
Let $b_1(v_i)$ be the bridge that precedes $p(v_i)$ 
and let  $b_2(v_i)$ be the bridge that 
follows $s(v_i)$; there are $O(d)$ elements between $b_1(v_i)$ and $b_2(v_i)$ 
in $AC(v_{i+1})$. Bridges 
$b_1(v_i)$ and $b_2(v_i)$  can be found by storing proper and 
improper elements  
 of 
$AC(v_i)$ as proper and auxiliary elements in a union-split-find data 
structure. 
Unfortunately, we would need $\Omega(\log \log n)$ time to identify 
$b_1(v_i)$ because of the lower bound of~\cite{MNA88}. 

Our solution does not rely on bridges and  union-split-find data structures 
 during the search procedure. Instead, 
we construct  additional catalogs $\oAC(u)$ in each node $u$. 
Catalogs $\oAC(v)$ are constructed in a leaf-to-root order: for a leaf 
node $u_l$, $\oAC(u_l)=AC(u_l)$; for an internal node $u$, $\oAC(u)$ 
contains all elements of $AC(u)$ and some elements from 
catalogs $\oAC(u_j)$, where $u_j$ are the  children  of $u$. 
We guarantee that at least one element from  a sequence of $\log^{O(1)}n$ 
consecutive elements in $\oAC(u)$ also belongs to $\oAC(\parent(u))$ 
for any node $u$. This allows us to identify the elements 
$b_1(v_i)$ and $b_2(v_i)$ that precede $p(v_i)$ and  
follow $s(v_i)$ in  $\oAC(v_i)\cap \oAC(v_{i+1}))$  in $O(1)$ time. 
Hence, we can quickly navigate from a node to its parent. 
On the other hand, 
 a catalog $\oAC(v_i)$, $i\geq 1$, can contain a large number of elements 
that are not relevant for the search procedure, i.e., elements 
from  some catalogs $C(w)$, such that $w$ is a descendant of $v_i$ 
and $w\not\in \pi$. 
Hence, there can be an arbitrarily large number of elements in 
$\oAC(v_i)\cap [b_1(v_i),b_2(v_i)]$. 
However, we can show that the number of elements in 
$AC(v_i)\cap [b_1(v_i),b_2(v_i)]$ is bounded.  

We store a data structure $R(v)$ in every node $v$ of $T$. 
For any two elements $e_1\in \oAC(v)$ and $e_2\in \oAC(v)$,
the data structure $R(v)$ identifies 
an element $e_3\in AC(v)$ such that $e_1\leq e_3 \leq e_2$, or determines 
that such $e_3$ does not exist. The data structure $R(v)$ combines 
the approach of the dynamic range reporting data structure~\cite{MPP05, M06} 
and the labeling technique~\cite{IKR81,W92}; details can be found in 
section~\ref{sec:mcs}. 
Using $R(v)$, we can identify relevant elements in every visited node 
on the search path $\pi$.

\section{Multiple Catalog Searching}
\label{sec:mcs}
{\bf Overview.}
In this section we give a more detailed description of the approach sketched 
in section~\ref{sec:overview}. Every node $v$ contains a catalog $C(v)$ 
and an augmented catalog $AC(v)\supset C(v)$. 
For the root $v_R$, $C(v_R)=AC(v_R)$. The augmented catalog for a non-root 
node $v$ contains some elements from catalogs $AC(w)$ for ancestors $w$ 
of $v$, so that the following statement is true:  
\begin{property}\label{prop:fc}
Let  $u_i$ be a child of an internal node $u$.
Suppose that for  two  elements $e_1\in AC(u)$ and $e_2\in AC(u)$ 
 there is no $e_3 \in  AC(u)\cap AC(\parent(u))$ such that  $e_1 < e_3 < e_2$. 
Then there are $O(d)$ elements $e\in AC(\parent(u))$, such that 
$e_1\leq e \leq e_2$.
\end{property}
Here and further $d$ denotes the maximal node degree of $T$. 
Using the standard fractional cascading technique~\cite{MN90,R92}, we 
can construct and maintain catalogs $AC(v)$ that satisfy 
Property~\ref{prop:fc}. 

\begin{figure}[tb]
\centering
\includegraphics[width=.8\textwidth]{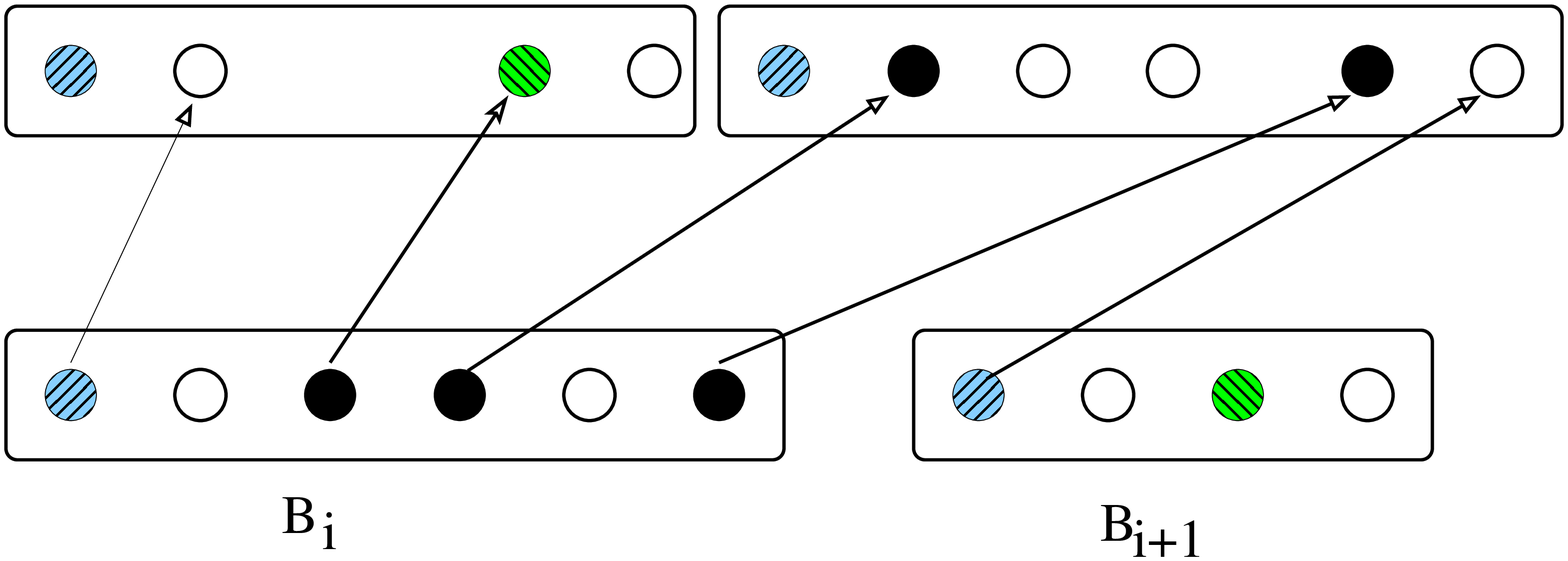}
\caption{\label{fig:oacstruct} 
An example of blocks in catalogs $\oAC(v)$ and $\oAC(\parent(v))$.
Elements $AC(v)\setminus C(v)$ and elements of $\tAC(v)$ are 
depicted with black circles and light blue circles respectively. Proper elements of $AC(v)$ are shown with green circles and all other 
elements of $\oAC(v)$ are shown with white circles.}
\end{figure}

Each node $v$ also contains a catalog $\oAC(v)\supset AC(v)$. 
Each catalog $\oAC(v)$ is subdivided into blocks $B_i$, so that 
(1) any element in a block $B_i$ is smaller than any element in a block 
$B_j$ for $i<j$, and (2) each block, except of the last one,  contains more 
than $\log^3 n/2$ and less than $2\log^3 n$ elements; the last block contains
at most $2\log^3n$ elements. The set $\tAC(v)$ consists of first elements 
from each block $B_i$ of $\oAC(v)$. 
For a leaf node $v_l$, $\oAC(v_l)=AC(v_l)$.  For an internal node $u$,
$\oAC(u)=AC(u)\cup(\cup_i \tAC(u_i))$, where the union is taken over all 
children $u_i$ of $u$.  Thus $\oAC(v)\cap \oAC(\parent(v))= \tAC(v) \cup 
(AC(v)\setminus C(v))$. 
Each element of $\tAC(v)$ and each   element in  
 $AC(v)\setminus C(v)$ contains a pointer to its copy in 
$\oAC(\parent(v)) $ called \emph{the up-pointer}. 
We denote by $UP(v)$ the set  of all elements in $\oAC(v)$ that have 
up-pointers, i.e., $UP(v)=\oAC(v)\cap \oAC(\parent(v))$. 
Thus $UP(v)$ consists of first elements in every block of $\oAC(v)$ 
and improper elements from  $AC(v)$.
See Fig.~\ref{fig:oacstruct}. 
By a slight misuse of notation, we will sometimes denote the elements of 
$UP(v)$ as up-pointers.
For each block $B_i$ of every catalog $\oAC(v)$, we store a data 
structure $\cB_i$ that enables us to find for any $e\in B_i$ the largest 
element $f\ \in (B_i\cap UP(v))$ such that $f\leq e$. Such queries 
can be supported in $O(1)$ time because a block contains a poly-logarithmic 
number of elements. 

We store in each node $v$ a data structure 
$R(v)$ that enables us to find an element $e\in AC(v)$ between any two 
elements of $\oAC(v)$, or determine that such $e$ does not exist. 
The data structure 
The data structure $R(v)$ and the following Property play a key role 
in our construction
\begin{property}\label{prop:search1}
Let $b_1$ and $b_2$ be two  elements of $UP(v)$ such that 
there is no element $f\in AC(v)\cap UP(v)$ with $b_1 < f < b_2$.  
Then the catalog $AC(\parent(v))$
 contains  $O(d)$ elements $e$, such that 
$b_1\leq e \leq b_2$. 
\end{property}
\begin{proof}
Property~\ref{prop:search1} is a straightforward corollary of 
Property~\ref{prop:fc}:  
Let $e_1=\spred(b_1, AC(v))$ and $e_2=\ssucc(b_2,AC(v))$. 
Since $(AC(v)\cap AC(\parent(v)))\subset  UP(v)$,
$(AC(v)\cap AC(\parent(v)))\subset (AC(v)\cap UP(v))$. 
Hence, there is no element of $AC(v)\cap AC(\parent(v))$ between 
$e_1$ and $e_2$. Therefore, by Property~\ref{prop:fc}, $AC(\parent(v))$ 
contains $O(d)$ elements $e$, such that $e_1\leq e \leq e_2$.
Since $e_1\leq b_1 < b_2\leq e_2$, Property~\ref{prop:search1} is true.
\end{proof} 
We observe that Property~\ref{prop:search1} only bounds the number 
of elements in $AC(\parent(v))\cap [b_1,b_2]$. 
The number of elements in $\oAC(\parent(v)) \cap [b_1,b_2]$ 
can be arbitrarily large. 

In the next part of this section we show how multiple catalog searching 
queries can be answered if Property~\ref{prop:search1} is satisfied.  
Then, we describe the data structure for a block and the data structure 
$R(v)$. 
Finally,  we describe the update procedure and sketch the 
analysis of the space usage and the update time.
\\{\bf Search Procedure.}
Let  $v_0$ be a node of $T$ and let $\pi$ be the path from 
$v_0$ to the root $v_R$ of $T$.  
We will describe the procedure that identifies both the predecessor 
and the successor of $x$ in $\cup_{v\in\pi}C(v)$. 
In every node $v\in \pi$ we identify  elements  
$p(v)=\spred(x,P_v)$ and $s(v)=\ssucc(x,P_v)$,
where $P_v=\cup_{u\in \pi_v} AC(u)$ and  $\pi_v$ is the path from $v_0$ to $v$.
Clearly, we can find $p(v_0)$ and $s(v_0)$ in time $O(t(n))$. 
Let $b_1(v_0)$ and $b_2(v_0)$ be the up-pointers that precede and follow 
$p(v_0)$ and $s(v_0)$. Since each block contains at least one up-pointer,
we can find $b_1(v_0)$ and $b_2(v_0)$ in $O(1)$ 
time. 

Suppose that we know $p(v_i)$, $s(v_i)$, $b_1(v_i)$, and $b_2(v_i)$ 
for some  node $v_i\in \pi$, so that $b_1(v_i)\leq x \leq b_2(v_i)$
and  $b_1(v_i)$, $b_2(v_i)$ are up-pointers 
that satisfy the condition of Property~\ref{prop:search1}. We can 
find $p(v_{i+1})$, $s(v_{i+1})$, $b_1(v_{i+1})$, and $b_2(v_{i+1})$
for the parent $v_{i+1}$ of   $v_i$  as follows. 
By Property~\ref{prop:search1}, there are at most $O(d)$ elements of 
$AC(v_{i+1})$ between $b_1(v_i)$ and $b_2(v_i)$. 
Since $b_1(v_i)\leq p(v_i) \leq s(v_i) \leq b_2(v_i)$, elements 
of $AC(v_{i+1})$ that do not belong to the interval $[b_1(v_i),b_2(v_i)]$ 
are not relevant for our search.  
If $AC(v_i)\cap [b_1(v_i),b_2(v_i)]\not=\emptyset$, we can identify 
some $e\in AC(v_{i+1})$, $b_1(v_i)\leq e\leq b_2(v_i)$, using the 
data structure $R(v_{i+1})$. 
We will show in the next paragraph how  all elements in 
$AC(v_{i+1})\cap [b_1(v_i),b_2(v_i)]$ can be examined and compared with $x$, $p(v_i)$, and $s(v_i)$ in $O(1)$ time. Hence, we can identify $p(v_{i+1})$ and $s(v_{i+1})$ 
in time $O(1)$. The up-pointers $b_1(v_{i+1})$ and $b_2(v_{i+1})$ 
are the up-pointers that precede $p(v_{i+1})$ in $AC(v_{i+1})$ 
and follow $s(v_{i+1})$ in $AC(v_{i+1})$ respectively.  
Otherwise, if $AC(v_i)\cap [b_1(v),b_2(v)]=\emptyset$, 
$p(v_{i+1})=p(v_i)$ and $s(v_{i+1})=s(v_i)$. 
In this case the up-pointers $b_1(v_{i+1})$ and $b_2(v_{i+1})$ are the 
up-pointers that precede  $b_1(v_{i})$ and follow $b_2(v_{i})$
respectively.
 Since every element in $AC(v)$ belongs either to $C(v)$ or 
to $C(w)$ for some ancestor $w$ of $v$, 
$\cup_{v\in\pi} AC(v)=\cup_{v\in\pi} C(v)$. 
Hence if we know $p_{v_R}$ and $s_{v_R}$ for the root node $v_R$, we also know
$\spred(x, \cup_{v\in \pi} C(v))=p_{v_R}$ and $\ssucc(x,\cup_{v\in \pi} C(v) )=s_{v_R}$.  

It remains to show how we can find  $p(v_{i+1})$ and $s(v_{i+1})$ 
if $AC(v_i)\cap [b_1(v_i),b_2(v_i)]\not=\emptyset$ and 
a pointer to some element  $e\in AC(v_{i+1})$, $b_1(v_i)\leq e\leq b_2(v_i)$,
is given. Suppose  that the maximal node degree $d=O(\log^g n)$ for 
a constant $g$.
We divide $AC(v)$ for each $v\in T$  into groups 
$G_j$ so that each group contains 
at least $\log^g n$ and at most $4\log^g n$  
elements and store the elements of each group in 
the atomic heap  $Q_i$ of Fredman and Willard~\cite{FW94}, 
so that predecessor queries and updates are supported in $O(1)$ time~\cite{FW94,T99}. 
There are $O(1)$ groups $G_j$, such that 
 $G_j\cap [b_1(v_i), b_2(v_i)]\not=\emptyset$. Hence, we can find the largest 
index $f$, such that the first element in $G_f$ is smaller than 
$x$ in $O(1)$ time. 
Using $Q_f$, we find the predecessor $p_f$ of 
$x$ in $G_f$. If $p_f$ is larger than $p(v_i)$,
we set $p(v_{i+1})=p_f$; otherwise  $p(v_{i+1})=p(v_i)$.
Hence, we can find $p(v_{i+1})$ in $O(1)$ time.
We can find $s(v_{i+1})$ with  the symmetric procedure. 

Thus our search procedure  answers one query to a data structure $R(v)$ in 
every node $v\in \pi$. All other operations take $O(1)$ time per node.

\begin{figure}
\centering
\includegraphics[width=.8\textwidth]{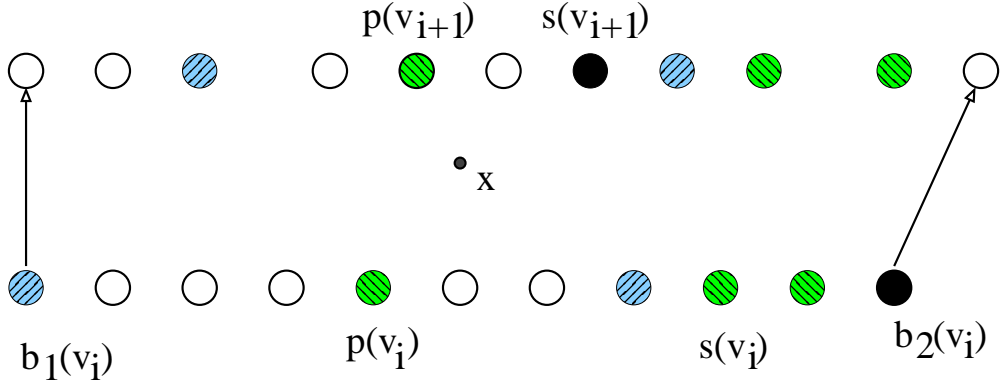}
\caption{\label{fig:search1} 
Searching for the predecessor and the successor of $x$ in a node $v_{i+1}$.
Elements $AC(v)\setminus C(v))$ and elements of $\tAC(v)$ are 
depicted with black circles and light blue circles respectively. 
Proper elements of $AC(v)$ are shown with green circles and all other 
elements of $\oAC(v)$ are depicted with white circles. Only relevant 
up-pointers are shown.}
\end{figure}

$~$\\{\bf Block Data Structure.}
Our data structure  uses the fact that a block contains $O(\log^3 n)$ elements.
Hence,  
each element of a block can be specified with  $O(\log\log n)$ bits 
and information 
about $\Theta(\sqrt{\log n})$ elements can be packed into one machine 
word. We can use this fact to store information about all elements of a 
block in a tree with node degree $\Theta(\sqrt{\log n})$. Details are 
given below. 
   
We associate a unique stamp $t(e)\leq 4\log^3 n$ with each element $e$ in $B$. 
The array $A$ contains entries for all elements of $B$ so that $A[k]=e$
for $t(e)=k$. We rebuild $A$ after $2\log^3 n$ update operations and 
assign an arbitrary stamp $t(e)\leq |B|$ to each $e\in B$. When a new element 
$e'$ is inserted into $B$, we set $t(e')=k'$, where $k'$ is the number of update operations since the last rebuild.

We also store a B-tree $T_B$ with node degree $\Theta(\sqrt{\log n})$ 
augmented as follows. Let $S(w_l)$ be the set of elements stored in 
a leaf $w_l$. The word $L(w_l)$ contains the time-stamps and ranks of all 
elements in $S(w_l)$.  
We also associate a word (i.e., a sequence of $O(\log n)$ bits) $M_w$ with
 each node $w$ of $T$. The $i$-th bit in
$M(w_l)$ for a leaf $w_l$ equals to $1$ if the $i$-th element of $S(w_l)$ 
belongs to $UP(v)$.
The $i$-th bit in $M(w)$ for an internal node $w$ equals to $1$ if and only if 
at least 
one bit in $M(w_i)$ equals to $1$, where $w_i$ is the $i$-th child of $w$.
For each word $M(w)$ and for any $i$, we can find the largest $j\leq i$, 
such that the $j$-th bit of $M(w)$ is set to $1$. 
Using a look-up table of size $o(n)$, common for all blocks, 
we can answer such queries in $O(1)$ time.
For each element $e$, we store a pointer to the leaf $w_l$ of $T_B$, 
such that $e$ belongs to $S(w_l)$.  

Given an element $e$, we identify the leaf $w_l$ in which it is stored. 
Using $L(w_l)$ we identify the rank  $j$ of $e$ in $S(w_l)$. 
This can be done in $O(1)$ time  with standard bit operations.
If there is at least one bit set to 1 among the first $j-1$ bits 
of $M(w_l)$, we use $L(w_l)$ to identify the stamp of the element $e'$ that 
corresponds to the $k$-th bit in $M(w_l)$, where $k$ is the index of the 
rightmost bit set to $1$ among the first $j-1$ bits of $M(w_l)$.  
Then, we find the element $e'$ using the array $A$.  
Otherwise, we search for the rightmost leaf 
$w_s$, such that $w_s$ is to the left of $w_l$ and $S(w_s)$ contains
at least one bit set to $1$.
Since the height of $T_B$ is $O(1)$, we can find $w_s$ in $O(1)$ time. 
Then, we use $L(w_s)$ and $A$ to identify the element corresponding 
to the rightmost bit set to $1$ in $w_s$. 

When  a new element is inserted, we insert an entry into the array $A$. 
Then, we identify the leaf $w_l$ in which $e$ is stored and update the 
word $L(w_l)$. We also update the word $M(w_l)$ and the words $M(w_j)$ 
for all ancestors $w_j$ of $w_l$. The B-tree can be re-balanced in a standard 
way. Deletions are performed symmetrically.
\\{\bf Data Structure $R(v)$.} 
Essentially, our data structure is based on the combination of 
the range reporting data structure of Mortensen, Pagh, and Patrascu~\cite{MPP05}
and the dynamic labeling scheme of~\cite{IKR81,W92}.  
Using the method of~\cite{W92}, we can assign a positive integer 
label bounded by $O(|\oAC(v)| / \log^3 n)$ to each block of $\oAC(v)$, 
so that 
labels can be inserted 
and deleted in $O(\log^2n)$ time. If a block contains at least one element
from $AC(v)$, then we store the label of this block 
in a data structure $R^u$ that supports one-dimensional range reporting 
queries.  Using the result of~\cite{MPP05}, the data structure $R^u$
supports queries  in $q_r(n)$ time and updates in time $u_r(n)$, 
where $q_r(n)$ and  $u_r(n)$ are arbitrary functions satisfying 
$u_r(n)\geq \log \log n$, $q_r(n)\leq \log \log \log n$, 
 and $2^{q_r(n)}=O(\log_{u_r(n)} \log n)$.  
For instance, queries and updates can be supported in 
$O(1)$ time and  $O(\log^{\eps}n)$ time respectively. 
Alternatively, $R^u$ can support queries in $O(\log\log \log n)$ time
and updates in $O(\log \log n)$ time. 
Although the data structure~\cite{MPP05} uses randomization and the update time
is expected, we can obtain the data structure with the same deterministic 
worst-case update time by replacing all Bloomier filters with bit vectors. 
The space usage of this modified data structure is $O(|\oAC(v)|/\log^2 n)$. 

We can determine, whether there is an element $e\in AC(v)$ between 
two elements $e_1$ and $e_2$ that belong to the same block $B_i$, using 
a data structure that is similar to the block data structure $\cB_i$. 

If $e_1$ and $e_2$ belong to different blocks $B_1$ and $B_2$, 
we can determine whether there is an element $e'$ such that 
$e'\in AC(v)\cap B_1$ and $e'$  is larger than $e_1$ 
or $e'\in AC(v)\cap B_2$ and $e'$ is smaller than $e_2$ as explained 
in the previous paragraph. If such $e'$ does not exist, we check 
whether there is a block between $B_1$ and $B_2$ that contains 
at least one element of $AC(v)$ using the data structure $R^u$.
If such a block $B_3$ is found, we identify an element $e'\in AC(v)\cap B_3$. 
\\{\bf Space Usage and Updates.}
It was shown in~\cite{MN90} that all catalogs $AC(v)$ contain $O(n)$ 
elements and an update on a catalog $C(v)$ incurs $O(1)$ amortized updates 
of catalogs $AC(u)$. An element $e$ can be inserted into or deleted 
from  a catalog $AC(u)$  in 
$O(\log \log n)$ time if the position of (the predecessor of) the 
element $e$ in $AC(u)$  is known: see e.g.,~\cite{MN90}. 
Applying the method of~\cite{MN90} to catalogs $AC(v)$, 
we can show that all catalogs $\oAC(v)$ 
also contain $O(n)$ elements, and an update of a catalog $AC(v)$ 
incurs $O(1)$ updates of $\oAC(w)$ for some nodes $w$. 

When a new element $e$ is inserted into $\oAC(v)$, we update the data 
structure for the block $B$ that contains $e$; we also update 
the data structure $R^u$ if $e\in AC(v)$. If the number of 
elements in a block equals to  $2\log^3 n$, we split the block into 
two blocks, so that each block contains $\log^3 n$ elements, 
insert a new label for one of the newly created blocks, 
and update the data structure $R^u$. 
When a new label is inserted, $O(\log^2 n)$ other labels may be changed. 
Hence, we must perform $O(\log^2 n)$ updates of the data structure 
$R^u$. Since a new label is inserted after $O(\log^3 n)$ insertions, 
the amortized  cost of an insertion into $R(v)$ is $O(u_r)$. 
If $e$ also belongs to $AC(v)$ and $e$ is the only element in 
$e\in AC(v)\cap B$, then $e$ must be inserted into 
a data structure $Q_j$ for some group $G_j$. If the number of elements 
in $G_j$ equals to $4\log^g n$, we split the group into two groups 
of equal size. Thus the amortized cost of an insertion into 
$G_j$ is $O(1)$. Deletions are performed symmetrically. 
Since the update time is dominated by an update of the data structure 
$R^u$, the total cost of an update operation is $O(u_r(n))$.

During the search procedure, we  must answer one query to a data 
structure $R(v)$ in every node $v\in \pi$; all other operations 
can be performed in $O(1)$ time. Hence, a query can be answered 
in $O(t(n)+|\pi|q_r(n) )$ time. 
The result of this section is summed up in the following Theorem.
\begin{theorem}\label{theor:mcs}
We are given a tree $T$ with maximal node degree $d=\log^{O(1)}n$, 
so that a catalog 
$C(v)\subset U$ is associated with each node $v$, $\sum_{v\in T} |C(v)|=n$.
Let $u_r(n)$ and $q_r(n)$ be arbitrary functions satisfying 
$u_r(n)\geq \log \log n$, $q_r(n)\leq \log \log \log n$, 
 and $2^{q_r(n)}=O(\log_{u_r(n)} \log n)$.  
There exists a data structure that 
answers multiple catalog searching queries in $O(t(n)+|\pi|q_r(n))$ time, 
where $t(n)$ denotes the time needed to search in one catalog of $n$ 
elements.
If a pointer to the (predecessor of) $x$ in $AC(v)$ is given, 
then $x$ can be inserted or deleted in 
$O(\log \log n + u_r(n))$ amortized time.
\end{theorem}
Two interesting choices of $u_r(n)$ and $q_r(n)$ are 
$u_r(n)=\log^{\eps} n$, $q_r(n)=\idtt{const}$ and $u_r(n)=\log \log  n$, 
$q_r(n)=\log \log \log n$. 
Thus  we can answer 
multiple catalog searching queries in $O(t(n)+|\pi|)$ time 
and support updates in $O(\log^{\eps}n)$ amortized time.
We can also answer 
multiple catalog searching queries in $O(t(n)+|\pi|\log \log \log n)$ 
time and support updates in $O(\log\log n)$ amortized time. 
\subsection{Multiple Catalog Reporting}
\label{sec:mcrep}
In this subsection we describe how our data structure can be modified 
to report  elements in the query interval $[x_l,x_h]$ for 
all catalogs $C(v)$, where $v$ is a node on a path $\pi$. In this case 
$\pi$ is a path  from a node $v_0$ 
to a node $v_1$ such that $v_1$ is the ancestor of $v_0$. 
We observe that, unlike in the multiple catalog searching problem, 
$v_1$ is not necessarily the root of $T$. 
\begin{theorem}\label{theor:mcrep}
We are given a tree $T$ with maximal node degree $d=\log^{O(1)}n$, 
so that a catalog 
$C(v)\subset U$ is associated with each node $v$, $\sum_{v\in T} |C(v)|=n$. 
Let $u_r(n)$ and $q_r(n)$ be arbitrary functions satisfying 
$u_r(n)\geq \log \log n$, $q_r(n)\leq \log \log \log n$, 
 and $2^{q_r(n)}=O(\log_{u_r(n)} \log n)$.  
There exists a data structure that 
answers multiple catalog reporting queries in $O(t(n)+|\pi|q_r(n) +k)$ time, 
where $t(n)$ denotes the time needed to search in one catalog of $n$ 
elements and $k$ is the number of points in the answer.
If a pointer to the (predecessor of) $x$ in $AC(v)$ is given, 
then $x$ can be inserted or deleted in 
$O(\log \log n + u_r(n))$ amortized time.
\end{theorem}
 We maintain the catalog $AC(v)$, the catalog $\oAC(v)$, and the data 
structure $R(v)$ in every 
node $v\in T$ as described in section~\ref{sec:mcs}. 
Moreover, every node $v$ contains
 a data structure $R_c(v)$: for any two elements $e_1$ and $e_2$ 
in $\oAC(v)$, $R_c(v)$ identifies an element $e'\in C(v)$, 
$e_1 \leq e' \leq e_2$, if such $e'$ exists. $R_c(v)$ is implemented in the 
same way as $R(v)$. 
In every node $v$ on the path $\pi$, we identify $p_l(v)\in \oAC(v)$ and  
$s_l(v)\in \oAC(v)$ such that 
$p_l(v) \leq x_l\leq s_l(v)$ and there is no $e\in AC(v)$ with  
$p_l(v)\leq e \leq s_l(v)$. 
We also identify $p_h(v)\in \oAC(v)$ and $s_h(v)\in \oAC(v)$ such that 
$p_h(v) \leq x_h\leq s_h(v)$ and there is no $e\in AC(v)$ with  
$p_h(v)\leq e \leq s_h(v)$. 
For any $e \in AC(v)$, $e\in [x_l,x_h]$ if and only if 
$s_l(v) \leq e \leq p_h(v)$.

We set $p_l(v_0)=\spred(x_l,AC(v_0))$ and $s_l(v_0)=\ssucc(x_l,AC(v_0))$.
Given $p_l(v)$ and $s_l(v)$ for some node $v\in \pi$, we identify the 
up-pointers $b_1(v)=\spred(p_l(v),UP(v))$ and $b_2(v)=\ssucc(s_l(v),UP(v))$.
Up-pointers $b_1(v)$ and $b_2(v)$ satisfy Property~\ref{prop:search1}.
Hence for the parent $w$ of $v$, the catalog $AC(w)$ contains at most
$r$ elements between $b_1(v)$ and $b_2(v)$. We can search for an element 
$e'\in AC(w)$, $b_1(v)\leq e' \leq b_2(v)$ using the data structure $R(w)$.
If such $e'$ does not exist, we set $p_l(w)=b_1(v)$ and $s_l(w)=b_2(v)$. 
Otherwise we examine $O(d)$  neighbors of $e'$ in $AC(w)$ and 
find $p_l(w)$ and $s_l(w)$. 
We can identify $p_h(v)$ and $s_h(v)$ for all nodes $v\in \pi$ in the same way. 
 
Since $C(v)\subset AC(v)$, any element $e\in C(v)$ belongs to the interval
 $[x_l,x_h]$ if and only if $s_l(v) \leq  e \leq  p_h(v)$. 
If $C(v)\cap [s_l(v),p_h(v)]\not=\emptyset$, we can 
find some $e_m\in C(v)\cap [s_l(v),p_h(v)]$ using the data structure $R_c(v)$.
Then, we examine elements that follow $e_m$ in $C(v)$ until an element 
$e_h\in C(v)$, $e_h>x_h$, is found. We also examine elements that precede 
 $e_m$ in $C(v)$ until an element $e_l\in C(v)$, $e_l< x_l$ is found. 
Thus we can report all elements in $C(v)\cap [x_l,x_h]$ 
in $O(|C(v)\cap [x_l,x_h]|)$ time if $s_l(v)$ and $p_h(v)$ are known.
Update time and space usage are the same as in the catalog searching data
 structure.

\section{Multiple Catalog Maximum Queries}
\label{sec:mcmax}
In this section we describe a simple data structure that 
enables us to identify the maximum element in each catalog $C(v)$ 
for every node $v\in \pi$ on a query path $\pi$. Again, $\pi$ is a path  from 
a node $v_0$ to a node $v_1$ such that $v_1$ is the ancestor of $v_0$.
In this section we assume that the maximum node degree of a node is 
$d=O(\log^{1/8} n)$.  

Moreover, we can support extended update operations. 
An operation $\mins(e,f,l,v)$ inserts an element $e$ into catalogs 
$C(v_f),C(v_{f+1}),\ldots,C(v_l)$, where $v_f,\ldots,v_l$ are children 
of some node $v$. In this case we say that an element is associated 
with an interval $[f,l]$ in the node $v$. We assume that each element is 
associated with at most one interval in every node $v$ of $T$.
An operation 
$\mdel(e,v)$ deletes an element $e$ from 
 all  catalogs $C(v_f),C(v_{f+1}),\ldots,C(v_l)$, such that $e$ is associated 
with an interval $[f,l]$ in the node $v\in T$.  
\begin{theorem}\label{theor:mcmax}
We are given a tree $T$ with maximal node degree $d=\log^{1/8}n$, 
so that a catalog 
$C(v)\subset U$ is associated with each node $v$, $\sum_{v\in T} |C(v)|=n$. 
There exists a data structure that 
answers multiple catalog maxima queries in $O(t(n)+|\pi|)$ time, 
where $t(n)$ denotes the time needed to search in one catalog of $n$ 
elements.
If a pointer to (the predecessor of) $x$ in $\cup AC(v_i)$ is given, 
then $\mins(x,f,l,v)$ and $\mdel(x,v)$ are supported in 
$O(\log\log n)$ time and $O((\log\log n)^2)$ time respectively. 
\end{theorem}
All elements from a catalog $C(v_i)$ are stored in a data structure $D(v)$ for 
a  parent $v$ of $v_i$. Each element $e$ in $D(v)$ is associated 
with an interval $[f,l]$, $f\leq l$, such that $e$ is stored in all catalogs 
$C(v_f),\ldots,C(v_l)$. We implement $D(v)$ using the generalized 
union-split-find data structure described in Theorem 5.2 of~\cite{GK09}. 
This enables us to support the following operations: we can insert 
a new element $e$ associated with an interval $[e_f,e_l]$ into $D(v)$ 
in $O(\log \log n)$ time if the position of $e$ in $\cup C(v_i)$ is 
known. We can delete an element $e$ from $D(v)$ in $O(\log \log n)$ 
time. For any interval $[x_1,x_2]$, $1\leq x_1\leq x_2\leq d$,
 and any $q$ we can find the largest 
element $e\in D(v)$ such that $e\leq q$ and 
$[e_f,e_l]\cap [x_1,x_2]\not=\emptyset$.

We store a local tree $T(v)$ in every node $v$. Leaves of $T(v)$ correspond 
to children of $v$; $T(v)$ is a binary tree of height $O(\log \log n)$. 
We say that an element $e$ \emph{covers} a node $u$ of $T(v)$ 
if $e$ is stored in all catalogs $C(v_i)$ for all children $v_i$ of $v$.
We say that $e$ \emph{belongs} to a node $u$ of $T(v)$ if 
$e$ covers $u$ but $e$ does not cover the parent of $u$. 
The set $F_u(v)$ contains all elements that belong to  a node $u$ of $T(v)$. 
The data structure $M(v)$ contains maximal elements from every set $F_u(v)$.

We can use the fact that $M(v)$ contains $O(\log n)$ elements and implement 
it in one machine word, so that for any path $\pi(v)$ in $T(v)$ 
the maximum element $e\in \cup_{u\in \pi(v)} F_u(v)$ can be found in 
constant time. Updates of  $M(v)$ are also supported 
in constant time.  $M(v)$ is implemented as follows. 
Let $\max_u$ denote the maximum element in $F_u(v)$. 
The word $W(v)$ contains  the \emph{rank} of $\max_u$ in $M(v)$  for 
every node $u$ of $T(v)$ (nodes of $T(v)$ are stored in pre-order).  
Since ranks of all $\max_u$ fit into one machine word, we can modify 
the ranks of all elements in $M(v)$ in $O(1)$ time 
when $\max_u$ for some node $u$ of $T(v)$ is changed. 
Using table look-ups and bit operations on $W(v)$, we can 
also find the maximum in  $\cup_{u\in \pi(v)} F_u(v)$ for any path 
$\pi(v)$ in $T(v)$.

Using data structures $M(w)$ for the parent $w$ of $v$,  we can find 
the maximum element in $C(v)$ for any node $v$ in constant time.
We can identify the maximum element in a catalog $C(v)$
 by finding the maximum element among $\max_{u}$ for $u\in \pi(v,w)$. 
Here $\pi(v,w)$ denotes the path in $T(w)$ from the leaf that corresponds 
to $v$ to the root of $T(v)$.
Hence, we can find the maximum element in each $C(v)$ in $O(1)$ 
time using $M(v)$. 

When a new element $e$ is inserted into catalogs $C(v_f),C(v_{f+1}),\ldots, 
C(v_l)$, 
we insert $e$ into the data structure $D(v)$, where $v$ is the parent node 
of $v_f,\ldots, v_l$. We can find $O(\log\log n)$ nodes $u_1,\ldots u_s$ 
in $T(v)$, such that $e$ belongs to each $u_j$, $1\leq j\leq s$.
For every $u_j$, if $e> \max_{u_j}$ then 
we update the data structure $M(v)$. 
Hence, an operation  $\mins(e,f,l,v)$ takes $O(\log \log n)$ time. 
When an element $e$ is deleted from catalogs $C(v_f),C(v_{f+1}),\ldots, 
C(v_l)$, we check whether $e$ is stored as a maximum element $\max_u$ 
for some nodes $u$ in $M(v)$. For every such $u$, we find the 
largest element $e_u\leq e$ such that $e_u\in F_u(v)$.
Using the data 
structure $D(v)$, we can find $e_u$ in $O(\log \log n)$ time. 
When $e_u$ is found, we update $M(v)$ accordingly in $O(1)$ time. 
Finally, we delete $e$ from $D(v)$ in $O(\log\log n)$ time. 
Hence, $\mdel(e,v)$ takes $O((\log\log n)^2)$ time.

\section{Applications}
\label{sec:appl}
Applications in which we associate ordered sets with each 
node of a balanced tree are a frequent topic  in data structures.
In many cases we want to search in all catalogs that are associated 
with nodes on a specified root-to-leaf path. 
Since a root-to-leaf path in a balanced tree consists of $O(\log n)$ 
nodes and $t(n)=O(\log n)$, where $t(n)$ is the time we need to search 
in one catalog of $n$ elements, Theorem~\ref{theor:mcs} enables us 
to spend $O(1)$ time in each catalog.
If the node degree of a balanced tree is $\Theta(\log^c n)$ for a constant 
$c$, then a root-to-leaf path consists of $O(\log n/\log \log n)$ nodes.
Using fusion trees~\cite{FW93, FW94}, we can search in a single  catalog 
in $t(n)=O(\log n/\log \log n)$ time. 
Hence, Theorems~\ref{theor:mcs} and~\ref{theor:mcrep} enable us to spend
 $O(1)$ time in each catalog  even in the case when the node degree is 
poly-logarithmic.  
Below we will sketch how our techniques can be used to obtain 
dynamic data structures for several important problems. 
\\{\bf Point Location in a Horizontal Subdivision.} 
In this problem the set of $n$ horizontal segments is stored in the 
data structure, so that for a query point $q=(q_x,q_y)$ the segment 
immediately below (or immediately above) $q$ can be reported. 
As in~\cite{GK09} and several other point location data
 structures~\cite{BJM94,ABG06}, our solution is based on 
segment trees. 
The leaves of a segment tree correspond to $x$-coordinates 
of segment endpoints. The range $rng(v)$ of a node $v$ 
is an interval $[p_l,p_r]$ where $p_l$ is the $x$-coordinate 
stored in the leftmost leaf descendant of $v$ and 
$p_r$ is the $x$-coordinate stored in  the rightmost leaf descendant of 
$v$. We denote by $proj(s)$ the projection of a segment $s$ on the $x$-axis. 
A set $S(v)$ is associated with each node $v$; 
$S(v)$ contains all segments $s$  such that $rng(v)$ is contained 
in $proj(s)$ but $rng(\parent(v))$ is not contained in $proj(s)$. 
Each internal node in our segment tree has $\Theta(\log^{\delta}n)$ 
children for $\delta=\eps/2$. Hence, each segment belongs to 
$O(\log^{1+\delta} n)$ sets $S(v)$.
If a $q_x\in proj(s)$ for some segment $s$, then $s$ is stored in one of sets 
$S(v)$, where $v\in \pi$ and $\pi$ is the  path from the leaf that contains 
the successor of  $q_x$ to the root of  the segment 
tree. 
We store the $y$-coordinates of segments from $S(v)$ in a catalog 
$C(v)$.
Hence finding a segment below (above) $q=(q_x,q_y)$ is equivalent 
to searching for the predecessor (successor) of $q_y$ in 
$\cup_{v\in \pi} C(v)$. 
We apply the multiple catalog searching technique to catalogs $C(v)$, 
so that we can search in $\cup_{v\in \pi} C(v)$ in $O(\log n/\log \log n)$ 
time and update a catalog $C(v)$ in $O(\log^{\delta}n)$ time. 
When a new segment $S$ is inserted into the data structure, we 
insert the $y$-coordinate of $s$ into $O(\log^{1+\delta} n)$ catalogs 
$C(v_1),\ldots,C(v_m)$.  
Using the standard fractional cascading technique, we can identify 
position of the $y$-coordinate $y_s$ of $s$ in 
augmented catalogs $AC(v_1),\ldots, AC(v_m)$ in
 $O(\log^{1+\delta}\log\log n)$
time. Then we can insert $y_s$ into the multiple catalog searching data 
structure  in 
$O(\log^{\delta}n\log^{1+\delta}n)=O(\log^{1+\eps}n)$ by 
Theorem~\ref{theor:mcs}.
Deletions are supported in the same way. 
The data structure uses $O(n\log^{1+\eps}n)$ space. But we can reduce the 
space usage to linear by using the technique described in~\cite{GK09} and 
the technique of~\cite{BJM94}. Details will be given in the full version of 
this paper.
\\{\bf Stabbing-Max Data Structure.}
We use the same construction as in the point location data structure, 
but catalogs $C(v)$ contain priorities of segments stored in $S(v)$. 
For this problem, we use Theorem~\ref{theor:mcmax}. 
To find the segment $s$ with the highest priority such that $x\in s$,  
we identify  the maximum element in catalogs $\cup_{v\in \pi} C(v)$.
Suppose that a new segment $s$ is inserted. All nodes $v$, such that 
$s$ is stored in $S(v)$ can be divided into $O(\log n/\log\log n)$ 
groups. The $i$-th group consists of sibling nodes $u_{i,f_i},\ldots, u_{i,l_i}$ that have the same parent node $u_i$. Using standard fractional cascading, 
we can identify the position of $s$ in all data structures $D(u_i)$ 
in $O(\log n)$ time. 
Then, we can use Theorem~\ref{theor:mcmax} to insert the segment $s$ into 
$D(u_i)$ and to update $M(u_i)$ in $O(\log \log n)$ time for each $u_i$. 
Hence, the total time for an insertion is 
$O(\log n)$. Deletions are performed in a symmetric way, but we 
need $O((\log \log n)^2)$ time to update $M(u_i)$. Hence, the total time 
for a deletion is $O(\log n\log\log n)$. We observe that it is not necessary 
to store catalogs $C(v)$ and set $S(v)$ in every node $v$. We only 
need to store the data structures $D(v)$ and $M(v)$ described in the proof 
of Theorem~\ref{theor:mcmax}. Hence, the total space used by our construction 
is $O(n\log n/\log \log n)$. Thus we obtain a $O(n\log n/\log \log n)$ space 
data structure that answers stabbing-max queries in $O(\log n/\log\log n)$ 
time, supports insertions in $O(\log n)$ time, and supports deletions in
$O(\log n\log\log n)$ time.

Alternatively, we can store $C(v)$  using Theorem~\ref{theor:mcs}.
To identify the highest priority segment that contains a query point $q$, 
we search for the predecessor of $p_{d}$ in $\cup_{v\in \pi} C(v)$; 
here $p_d$ denotes the dummy priority such that $p_d$ is larger than 
priority of any segment in the data structure. 
In this case update time and space usage can be estimated as for  the 
horizontal point location data structure.
Thus we obtain a $O(n)$ space data structure that supports queries in $O(\log n/\log\log n)$ time and updates in $O(\log^{1+\eps}n)$ time.
\\{\bf Line-Segment Intersection.}
Again, we use the same construction as in the point location data
 structure. But now we use Theorem~\ref{theor:mcrep}, so that multiple catalog 
 reporting  queries can be answered.
Given a vertical segment $s_q$ with endpoints $(x_q,y_1)$ and 
$(x_q,y_2)$,  each  segment that intersect $s_q$ belongs to some 
set $S(v)$ for a node $v\in \pi$, where $\pi$ is the path from the 
node that contains the successor of $x_q$ to the root of the segment 
tree. A segment $s\in S(v)$, $v\in \pi$, intersects with $s_q$ if and only if 
the $y$-coordinate of $s$ belongs to the range $[y_1,y_2]$. 
Hence, we can find all segments that intersect $s_q$ by answering a 
multiple catalog reporting query.
As in the previous case the update time is $O(\log^{1+\eps}n)$. 
We need $O(n\log^{1+\eps}n )$ space to store all segments.
 But we can reduce the space usage to 
$O(n\log n/\log \log n)$ by using the technique similar to 
the compact representation described in~\cite{B08}. Details will be given in
 the full version.

\end{document}